\documentclass[10pt,twocolumn,letterpaper]{article}

\usepackage{cvpr}   
\usepackage[symbol]{footmisc}

\usepackage{url}

\definecolor{cvprblue}{rgb}{0.21,0.49,0.74}
\usepackage[pagebackref,breaklinks,colorlinks,allcolors=cvprblue]{hyperref}
\usepackage[accsupp]{axessibility}  

\def\paperID{7628} 
\def\confName{CVPR}
\def\confYear{2025}

\title{RASP: Revisiting 3D Anamorphic Art for Shadow-Guided Packing of \\ Irregular Objects}

\author{
Soumyaratna Debnath\textsuperscript{1}\footnotemark[1] \quad
Ashish Tiwari\textsuperscript{1}\footnotemark[1] \quad
Kaustubh Sadekar \textsuperscript{2} \quad
Shanmuganathan Raman\textsuperscript{1} \\
\textsuperscript{1}Indian Institute of Technology Gandhinagar \quad
\textsuperscript{2}Portland State University \\
{\tt\small \{debnathsoumyaratna, ashish.tiwari, shanmuga\}@iitgn.ac.in, ksadekar@pdx.edu}
}

\begin{document}
\twocolumn[{%
\renewcommand\twocolumn[1][]{#1}%
\maketitle
\vspace{-0.75cm}
\begin{center}
    \centering
    \captionsetup{type=figure}
    \includegraphics[width=1\textwidth]{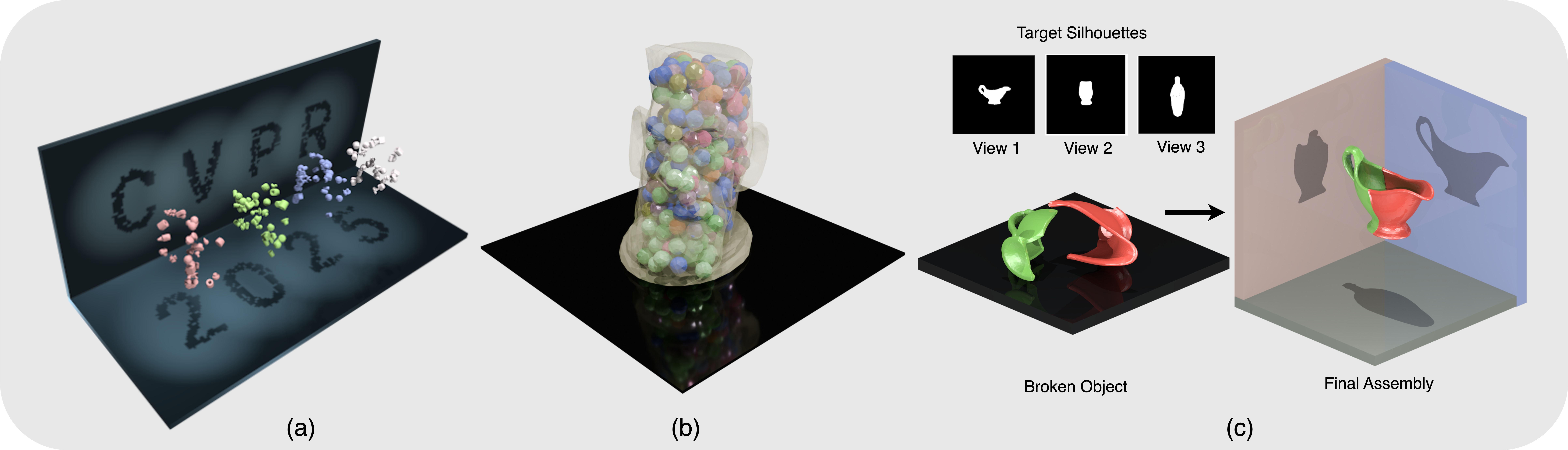}
    \captionof{figure}{(a) An ensemble of arbitrary shapes casting the shadow of alphabets CVPR and numbers 2025, (b) a set of irregular objects packed inside a face-shaped container, (c) an assembly of parts of a vessel obtained through multi-view shadow guidance via RASP.}
    \label{fig:teaser}
\end{center}%
}]

\footnotetext[1]{Equal Contribution $|$ \href{https://soumyaratnadebnath.github.io/RASP/}{Project Page}}

\begin{abstract}
Recent advancements in learning-based methods have opened new avenues for exploring and interpreting art forms, such as shadow art, origami, and sketch art, through computational models. One notable visual art form is 3D Anamorphic Art in which an ensemble of arbitrarily shaped 3D objects creates a realistic and meaningful expression when observed from a particular viewpoint and loses its coherence over the other viewpoints. In this work, we build on insights from 3D Anamorphic Art to perform 3D object arrangement. We introduce RASP, a differentiable-rendering-based framework to arrange arbitrarily shaped 3D objects within a bounded volume via shadow (or silhouette)-guided optimization with an aim of minimal inter-object spacing and near-maximal occupancy. Furthermore, we propose a novel SDF-based formulation to handle inter-object intersection and container extrusion. We demonstrate that RASP can be extended to part assembly alongside object packing considering 3D objects to be ``parts" of another 3D object. Finally, we present artistic illustrations of multi-view anamorphic art, achieving meaningful expressions from multiple viewpoints within a single ensemble.
\end{abstract}

\section{Introduction}
\label{sec:intro}
For centuries, artists have used their expressions to redefine the boundaries of visual art, demonstrating how art shapes reality and influences human perception. Their work has also broadly impacted technology, design, and engineering. In this work, we explore using a unique visual art form, 3D Anamorphic Art, to tackle challenges in irregular object packing and extend this approach to part assembly. 

Interestingly, we are not alone in using (and, in fact, developing) artistic expressions to address applications in computer vision. For example, DeepDream \cite{mordvintsev2015inceptionism} uses GAN-based style transfer to apply the style of classical paintings, like those of Van Gogh or Picasso, to contemporary images. RePaint \cite{shi2018deep} leverages deep learning and 3D printing to replicate the colors and textures of paintings by optimizing ink layering. ScribGen \cite{debnath2024scribgen} creates human-like scribbles and hand-strokes in different styles. Beyond images, researchers have expanded artistic techniques into 3D analysis and understanding, as in Shadow Art \cite{mitra2009shadow}, which uses shadows for 3D reconstruction, sketch-based 3D generation \cite{sadekar2022shadow}, expressive hand movements \cite{gangopadhyay2023hand}, and the design of knot configurations for realistic visual effects \cite{gangopadhyay2024search}. In this work, we draw upon insights from 3D Anamorphic Art to address an important problem of packing arbitrarily shaped objects within a 3D bounding volume (or container). Packing has numerous practical applications, spanning fields such as combinatorial optimization \cite{martello2000three, seiden2002online}, computational geometry \cite{hu2020tap,ma2018packing}, computer vision and graphics \cite{liu2019atlas, chen2015dapper,saakes2013paccam}, machine learning \cite{hu2017solving,zhao2021learning}, robotics \cite{yang2021packerbot,shome2019towards,wang2020robot}, and logistics and manufacturing \cite{zhao2023learning}.

\begin{figure}[t]
    \centering
    \includegraphics[width=\linewidth]{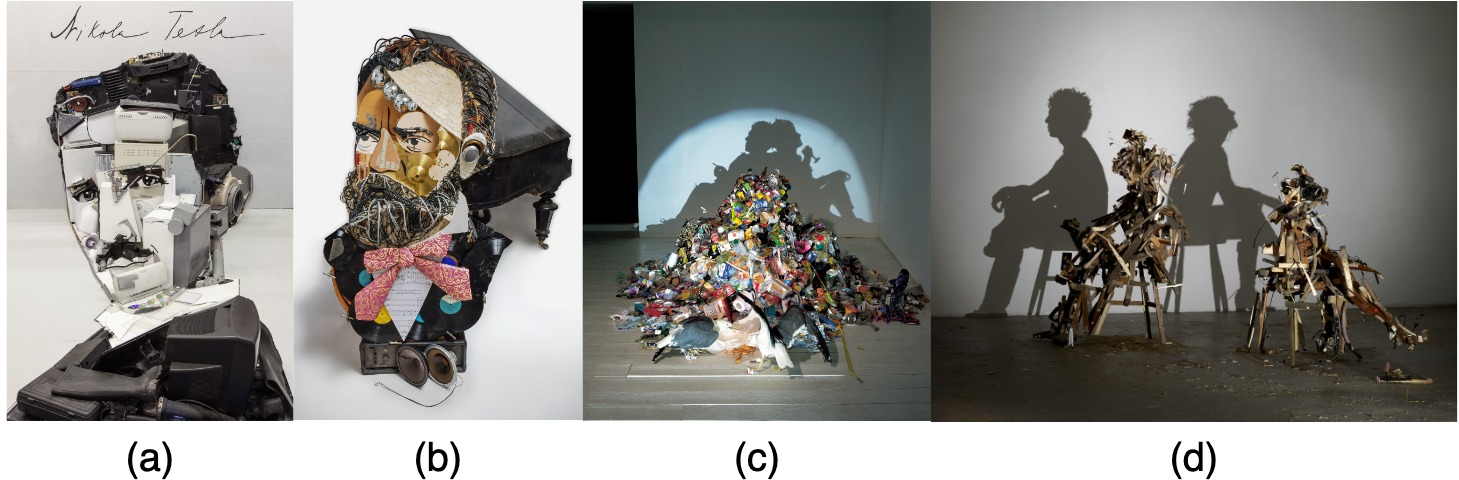}
    \caption{Illustration of 3D Anamorphic Art (a, b): Portrait of Nikola Tesla and Bed\v{r}ich Smetana by Patrick Pro\v{s}ko and Shadow Art (c, d): \textit{Dirty White Trash} and \textit{Wild Mood Swing} by Tim Nobel and Sue Webster.}
    \label{fig:1}
\end{figure}

\textbf{3D Anamorphic Art.} 3D Anamorphic Art involves arranging objects in 3D space such that the arrangement appears coherent and meaningful only from the specific viewpoint(s). In contrast, from any other view, it appears random. For example, artist Patrick Pro\v{s}ko used this technique at the Illusion Art Museum Prague by arranging electrical appliances to form a portrait of Nikola Tesla \cite{ana_tesla} (Figure \ref{fig:1} (a)) and arranging musical instruments to create a likeness of the musician Bed\v{r}ich Smetana \cite{ana_smetana} (Figure \ref{fig:1} (b)). This concept also extends to shadows cast by such arrangements under particular camera-light configurations. For instance, artists Tim Noble and Sue Webster created \textit{Dirty White Trash} \cite{dirty_white_trash} and \textit{Wild Mood Swings} \cite{wild_mood_swings}, where carefully arranged trash and wooden pieces cast shadows of people sitting beside each other (see Figure \ref{fig:1} (c, d)). Whether through perspective views from specific viewpoints or shadows created under particular camera-light configurations, these projections reveal valuable insights about the arrangement of objects in 3D space.

\begin{figure}[t]
    \centering
    \includegraphics[width=\linewidth]{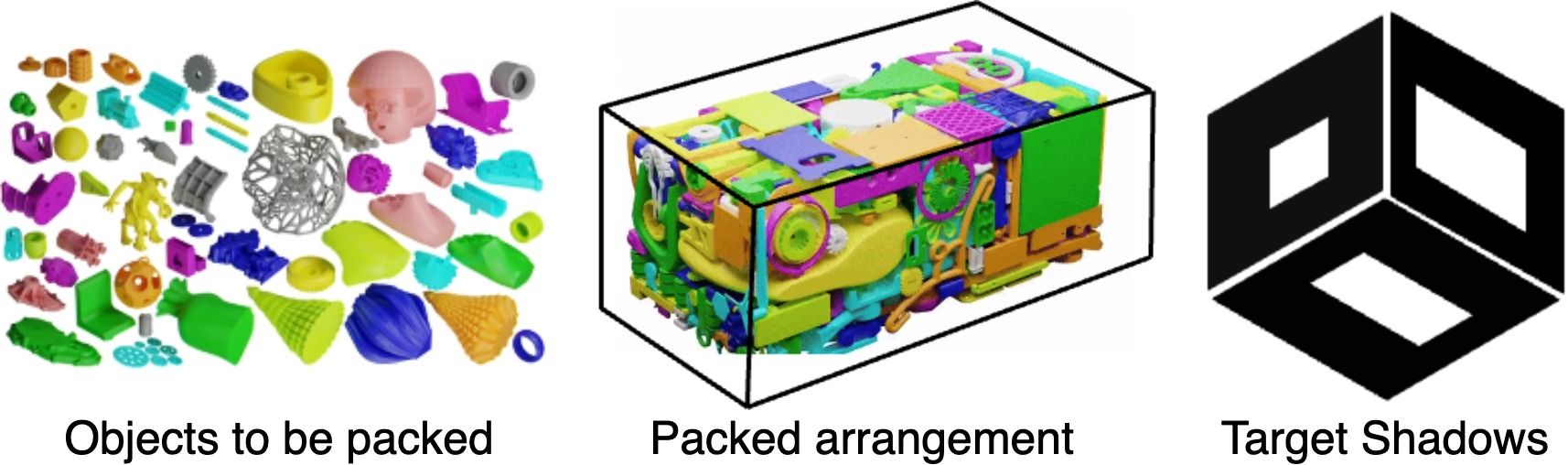}
    \caption{An example depicting the packed state of a cuboidal container and the associated silhouettes/shadows.}
    \label{fig:2}
\end{figure}

\textbf{Relation with Object Packing.} Consider a cuboidal bounding volume (or container), as depicted in Figure \ref{fig:2}. When fully packed, this cuboid would appear as a rectangle or square upon being viewed orthographically along any of its faces. Similarly, it would cast rectangular or square shadows under the same viewing configuration. Ideally, gaps between the objects would appear as holes in the shadow from at least one perspective. By considering these shadows (technically binary images or silhouettes of the packed state) as target projections, we aim to arrange arbitrarily shaped elements within a bounding volume, thereby tackling the packing problem.

Interestingly, this approach also allows for bounding volumes of arbitrary shapes, where the target projections are simply the volume's appearance from multiple viewpoints. If the elements to be arranged are whole 3D objects, this approach addresses the \textit{irregular object packing} problem. However, if these elements are ``parts" of another object altogether, it naturally extends to the \textit{part assembly} problem.

A key challenge in existing packing methods is managing overlaps or intersections among the elements to be packed. While some methods focus on simple geometries like cuboids \cite{crainic2012recent,yamazaki20003d}, spheres \cite{mackay1962dense}, cylinders \cite{stoyan2009packing}, or ellipsoids \cite{kallrath2017packing}, or rely on assumptions of convexity or concavity in polyhedrons \cite{stoyan2005packing,romanova2018packing,liu2015hape3d}, often using simple heuristics \cite{limper2018box, noll2011efficient}, others handle irregular shapes by employing voxelized representations \cite{byholm2009effective, de2013random}, which tend to be susceptible to approximation errors. This work introduces a signed-distance field (SDF) approach to manage inter-object intersection and container extrusion. 

\begin{figure*}[t]
    \centering
    \includegraphics[width=\textwidth]{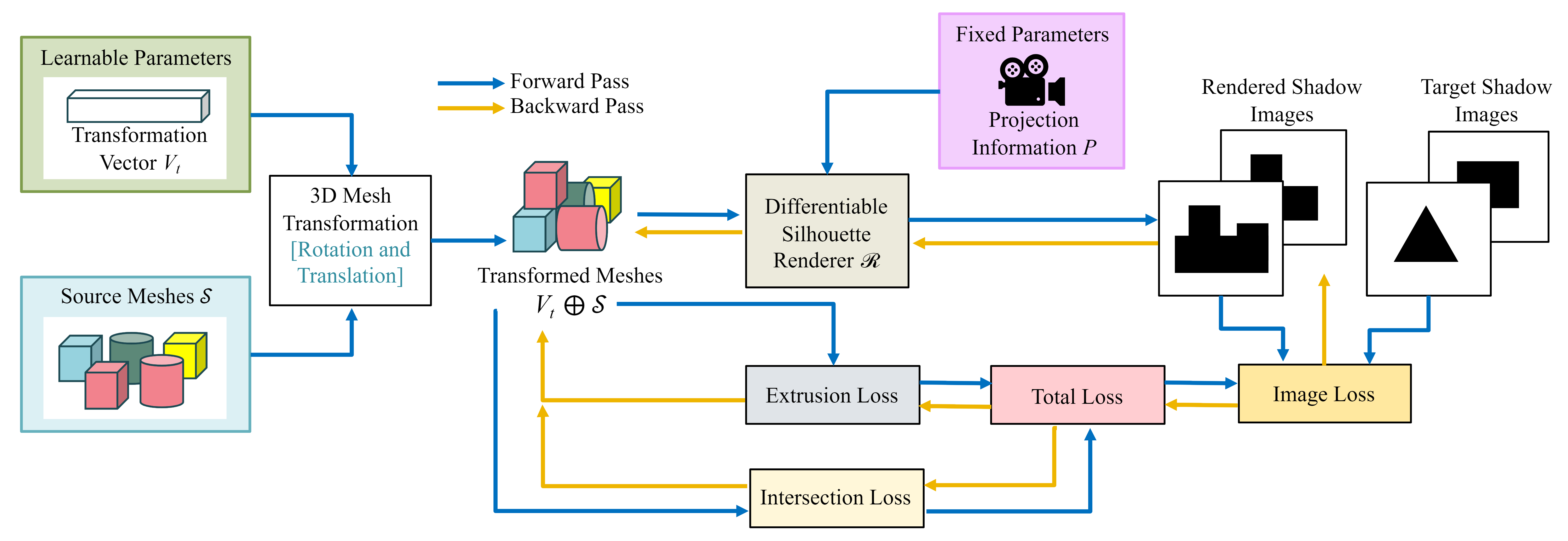}
    \caption{The proposed differentiable rendering-based pipeline for RASP. We use Adam optimizer with a learning rate of $1e-2$ to $1e-4$ over $1000$ iterations. Packing the Kitchen dataset ($106$ objects, $80000$ grid points) takes nearly $17$ minutes, and reassembly for each object in the Fantastic Breaks dataset ($2$ parts, $80000$  grid points) takes under $3$ minutes.}
    \label{fig:flow}
\end{figure*}

\textbf{Contributions.} In this work, we introduce RASP --\underline{\textbf{R}}evisiting 3D \underline{\textbf{A}}namorphic art for \underline{\textbf{S}}hadow-guided \underline{\textbf{P}}acking of Irregular Objects. Given a set of arbitrarily shaped objects, a bounding volume (or container), and information about the appearance of its projection/shadows from different viewpoints in the packed state:

\begin{itemize} \item We propose a differentiable rendering-based framework to tackle irregular object packing by drawing inspiration from 3D Anamorphic Art. Our goal is to achieve near-maximal occupancy and minimal inter-object spacing within a known bounding volume. 

\item We present a novel SDF-based approach to manage inter-object intersections and object-container extrusions, enhanced by an image-based loss function. 

\item We demonstrate that RASP can also be applied to part assembly without the need for explicit 3D ground truth supervision. Additionally, we illustrate compelling visual effects that cater to multi-view anamorphic art. 

\item To the best of our knowledge, this is the first approach to address packing (and part assembly) using only shadows or projections, guided by the principles of differentiable rendering.


\end{itemize}

Through this work, we push the boundaries of 3D anamorphic art to address an important research question: \textit{can shadows be used to automatically find an optimal arrangement of arbitrarily shaped 3D objects within a bounded volume?} While our framework aims to provide practical solutions, it does not always yield an optimal arrangement due to the NP-hard nature of the packing problem \cite{hartmanis1982computers}. Nonetheless, this work introduces a new perspective on solving the 3D packing and arrangement problem using shadows as a guiding mechanism.


\section{Related Works} \label{sec:related_work}
Object packing has been addressed using heuristic-based, model-based, learning-based, or policy-based (online) object packing. Several methods have assumed simple object geometries, such as cuboids \cite{crainic2012recent,yamazaki20003d}, spheres \cite{mackay1962dense}, cylinders \cite{stoyan2009packing}, or ellipsoids \cite{kallrath2017packing}, or leverage convexity and/or concavity in polyhedrons \cite{stoyan2005packing,romanova2018packing,liu2015hape3d}. This section primarily discusses the methods dealing with irregular object packing.

\textbf{Heuristic-based Methods.} Several previous attempts at object packing were primarily based on heuristic strategies. One method proposed by Wang \emph{et al.} \cite{wang2010two}  sequentially places objects employing the Deepest Bottom Left Fill (DBLF) heuristic strategy. However, it is shown to create holes, leading to empty spaces. Later, Wang and Hauser \cite{wang2021dense} addressed this limitation by reducing unfilled gaps through height-map minimization, which in itself ignores the difference between positions at the same level. HAPE3D \cite{liu2015hape3d}, another heuristic-based algorithm, deploys the principle of the minimum total potential energy for irregular polyhedrons. Even after allowing free translations and rotations, it fails to achieve high packing density. Lamas \emph{et al.} \cite{lamas2023voxel} voxelized irregular objects and used meta-heuristic algorithms to optimize voxel representation that is limited due to memory constraints. Heuristic-based methods generally apply well to objects with only a few facets and struggle to achieve a high packing density while dealing with complex irregular objects.

\textbf{Learning-Based Methods.} The key challenge for packing irregular objects is to handle object collision detection or intersection/overlap avoidance. Several different geometric representations of objects (say, objects tightly bound by a sphere or cuboid \cite{yao2015level}) have been assumed by different methods for collision detection and packing optimization. Romanova \emph{et al.} [1] pose the packing problem as a nonlinear optimization problem and propose quasi-phi functions for simple concave polyhedrons to describe non-overlapping and distance constraints. However, obtaining a local optimum takes longer, even with free translations and rotations. A few works \cite{chen2015dapper, yao2015level} learned to decompose the object into multiple small parts to perform packing to reduce the supporting materials, build time, and assembly costs of 3D printing. While \cite{chen2015dapper} used voxel-based representation, \cite{yao2015level} used level sets to represent the object parts.

\begin{figure*}[t]
    \centering
    \includegraphics[width=\linewidth]{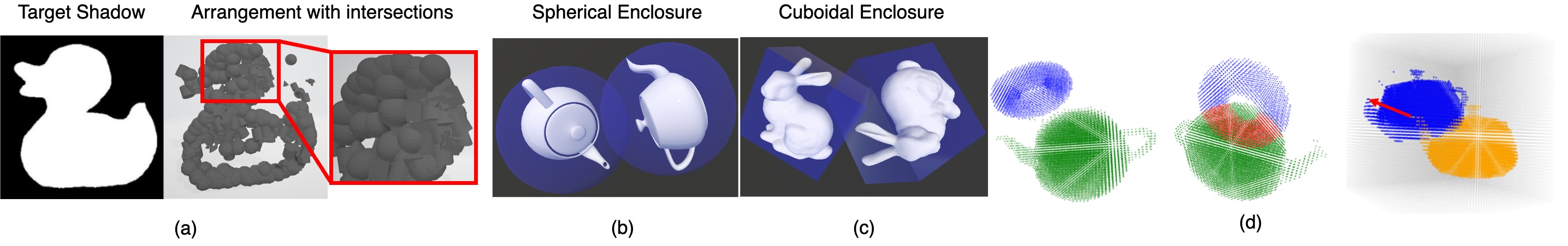}
    \caption{(a) Target shadow a duck-shaped container, optimized arrangement with mesh intersections. Mesh enclosed within (b) sphere and (c) cube where the intersection region doesn’t include object meshes. (d) SDF of two non-intersecting and intersecting objects (intersecting points are marked in red), and SDF transformation within the container.}
    \label{fig:intersection}
\end{figure*}

Moreover, Zhao \emph{et al.} \cite{zhao2023learning} and Zhuang \emph{et al.} \cite{zhuang2024dynamics} divide the objects into convex shapes for efficient collision detection in physical simulation and pack the objects either using shaking of the container as per the dynamic principle or reinforcement learning in an online manner, respectively. Our work does not focus on online packing. Cui \emph{et al.} \cite{cui2023dense} performed efficient collision metric computation in the spectral domain using voxel representation and Fast Fourier Transform (FFT) for discrete placement search. They deploy a greedy strategy to sort objects by volume from largest to smallest and then pack sequentially to minimize the distance between the new and already placed objects at each placement.

The closest to our setting is the work by Ma \emph{et al.} \cite{ma2018packing} that optimizes the orientation and position of the objects from their initial placement. However, their approach is multi-stage, involving combinatorial optimization that includes object swapping, enlargement, and replacement to reduce the gaps between the objects. Furthermore, they also minimize the height of a container to obtain an optimal one through binary search.  Due to repeated processing, the method is heavily time-consuming. In contrast, given a set of irregular objects, ours is a single-stage end-to-end optimization pipeline that performs packing only via 2D shadow guidance without any intense calculation within the 3D bounding volume. While Ma \emph{et al.} \cite{ma2018packing} have the information about the object set and process only one object at a time, sequentially determining the initial placement, our proposed method considers a set of objects at a time.

\section{Method}
\label{sec:method}
\subsection{Overview} 

The key idea of our work is to arrange a set $\mathcal{S}$ of arbitrarily shaped 3D objects within an arbitrarily shaped container $\mathcal{C}$ such that the resulting arrangement casts $K$ different shadows (projections) when viewed from $K$ different directions using a differentiable rendering-based optimization pipeline. Interestingly, the problem has two interpretations depending on how we consider the set $\mathcal{S}$. 

(a) \underline{\textit{3D Packing problem}}: when $\mathcal{S}$ contains arbitrarily shaped 3D objects and $\mathcal{C}$ is the bounding container within the viewing volume. 

(b) \underline{\textit{Part Assembly problem}}: when $\mathcal{S}$ contains ``parts" of a 3D object $(O)$ where this object $O$ itself is treated as the container $\mathcal{C}$. 

In the case of packing, we might not always have a ground truth or a unique solution, i.e., objects can be arranged in multiple optimal ways to maximize occupancy and minimize spaces. However, we have a ground truth shape in a part assembly whose parts can be assembled only in one way. We demonstrate how this is implemented in Section \ref{sec:exp}. Overall, we propose a differentiable-rendering-based framework that naturally caters to both these requirements, in addition to generating multi-view anamorphic art. Inspired by computational visual arts, some of the recent works based on shadow art \cite{mitra2009shadow, gangopadhyay2023hand, gangopadhyay2024search} have demonstrated the potential of 3D shape understanding, analysis, and reconstruction through shadows.  In this work, we take the first step towards leveraging the benefit of differentiable rendering and principles of visual arts \cite{wu2022survey} to propose shadow-based 3D packing and part assembly.

\subsection{Proposed Approach}

The flow of the proposed methodology has been outlined in Figure~\ref{fig:flow}. Consider a set of $N$ arbitrarily shaped objects $\mathcal{S}=\{{O}_{1},{O}_{2},…,{O}_{N}\}$ such that each ${O}_{i}$ is represented as a triangular mesh, an arbitrarily shaped bounding container $\mathcal{C}$ located inside the camera viewing volume $\mathcal{V}$, and shadow projection configuration $\mathcal{X} = \{X_{k}=(I_{k}, P_{k}) | k= 1,2,…, K\}$. Here, $\{I_{k}\}_{k=1}^{K}$ are the target images under the camera configuration $\{P_{k}\}_{k=1}^{K}$. Essentially, the camera configuration $P_{k}=(R_{k},t_{k})$ corresponds to the camera extrinsic parameters associated with image $I_{k}$. With $\mathcal{S}$, $\mathcal{C}$, and $\mathcal{X}$ as input to the differentiable renderer $\mathcal{R}$, the objective here is to learn the optimal arrangement of objects in set $\mathcal{S}$ inside the container $\mathcal{C}$ placed within the viewing volume $\mathcal{V}$ such that the resulting images $\{\widehat{I_{k}}\}_{k=1}^{K}$ rendered under camera configurations $\{P_{k}\}_{k=1}^{K}$ are close to the corresponding target images $\{I_{k}\}_{k=1}^{K}$. From the shadow’s perspective, the target images are silhouettes, and we assume that the light (to cast the shadows) will be co-located with the camera. Finding the arrangement of objects is equivalent to finding the rigid transformation, \emph{i.e.}, rotation and translation $\{R_{i},t_{i}\}_{i=1}^{N}$ of each object $O_{i}$ in set $\mathcal{S}$. We use quaternions for obtaining rotations - (a) to avoid gimble lock through other representations like axis-angle rotation or rotation composition of rotation along $x-y-z$ axes, (b) for smooth interpolation, and (c) to learn a smaller number of parameters compared to the rotation matrices. Specifically, we obtain the angle of rotations across each of the coordinate axes and convert them to quaternion representation for optimization.

\begin{figure*}[t]
    \centering
    \includegraphics[width=\textwidth]{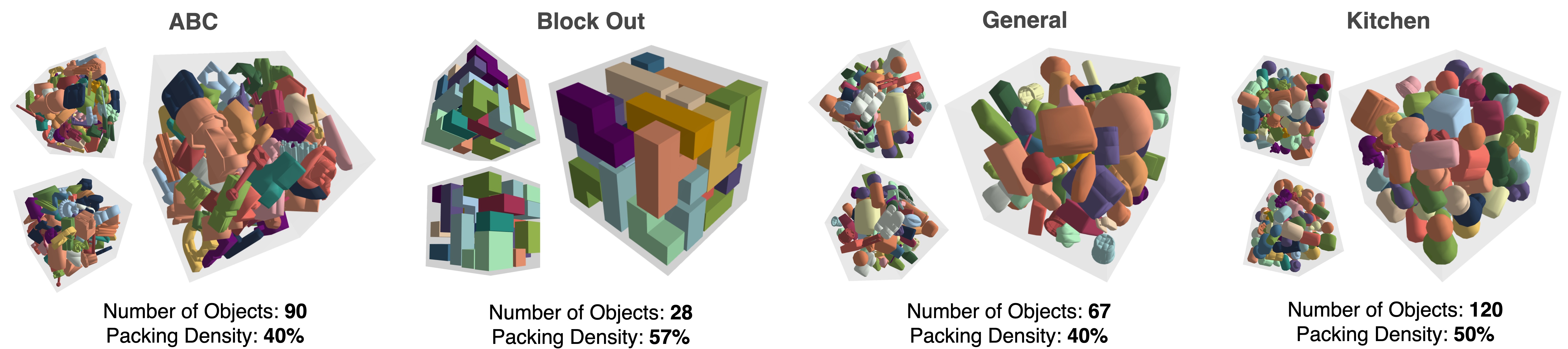}
    \caption{Visualization of 3D Object packing into a cuboidal container using RASP over four different object categories from \cite{zhao2023learning}. Each instance is labeled with the number of objects packed in the container and the packing density.}
    \label{fig:container}
\end{figure*}
\begin{figure}[h]
    \centering
    \includegraphics[width=\linewidth]{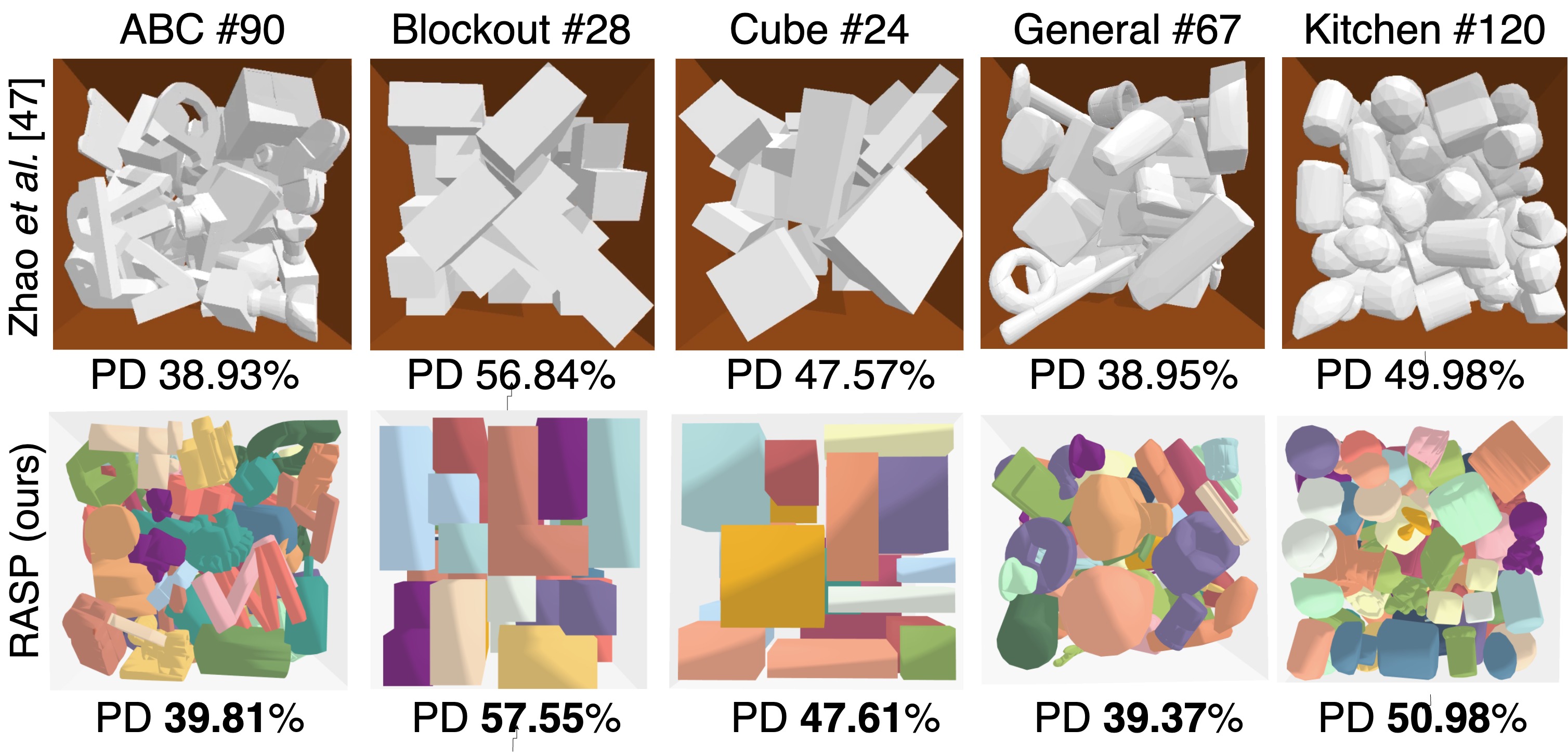}
    \caption{Qualitative comparison of object packing with RASP and Zhao \textit{et al.} \cite{zhao2023learning} for same set and same number of objects. Best viewed in PDF with zoom.}
    \label{fig:qual_res}
\end{figure}

\subsubsection{Objective Function}
We propose a combined loss function consisting of an SDF-based intersection loss, an image-based loss, and an object-container extrusion loss, as per Equation \ref{ref:total_loss}.
\begin{equation}
    \centering
    \mathcal{L}_{total} = \mathcal{L}_{sil} + \mathcal{L}_{is} + \lambda \mathcal{L}_{ext}
    \label{ref:total_loss}
\end{equation}
Here, we set $\lambda = 0.001$

\textbf{\textit{Image-based (Silhouette) Loss.}} A straightforward approach is to optimize each object's rotation and translation parameters based on image rendering loss, such as mean squared error, as per Equation \ref{eq:image_loss}. 
\begin{equation}
    \mathcal{L}_{sil} = \frac{1}{MNK}\sum_{k=1}^{K} \sum_{i=1}^{MN} ||I_{k}(i) - \widehat{I_{k}}(i)||_{2}^{2}
    \label{eq:image_loss}
    \end{equation}
Here, $MN$ is the total number of pixels in the image. As shown in Figure \ref{fig:intersection} (a), the projection of the arrangement may appear feasible (apart from a few gaps where more objects could fit), but in 3D, objects can still intersect within a duck-shaped container. 

\begin{figure}[!ht]
    \centering
    \includegraphics[width=\linewidth]{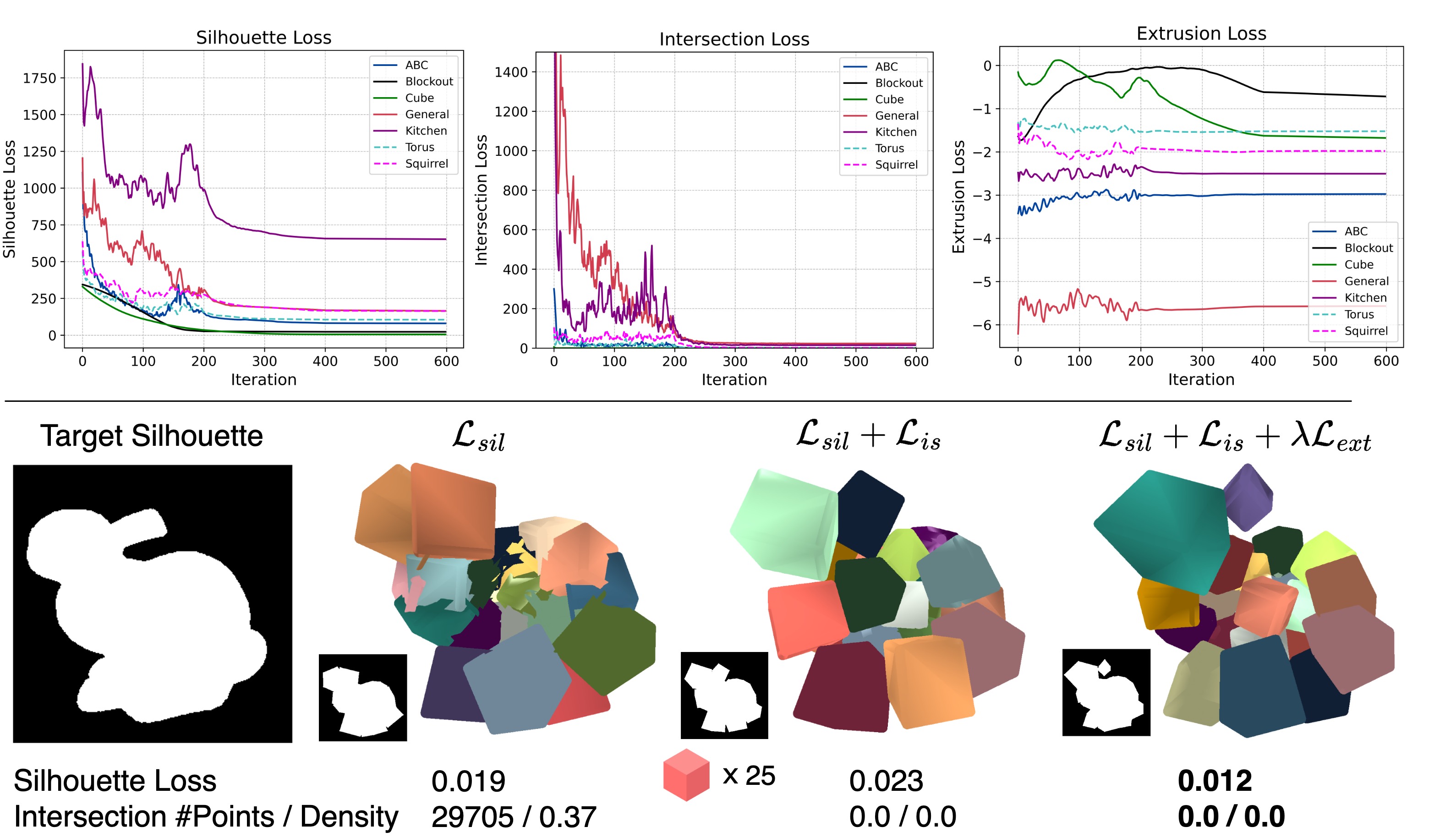}
    \caption{(Top) Loss curves for packed configuration in Figure \ref{fig:container} (solid line) \& Figure \ref{fig:custom_container} (dashed line). (Bottom) Effect of different loss terms on the packing. Best viewed in pdf with zoom.}
    \label{fig:ablation}
\end{figure}

Therefore, the loss functions should penalize mesh intersections. However, handling mesh intersections is complex; some methods \cite{karras2012maximizing} detect intersections by evaluating each triangular face of one mesh against others, which is computationally costly and impractical for large numbers of meshes. To simplify intersection computation, some approaches enclose objects within unit spheres or cubes, finding intersections among these simpler bounding shapes \cite{yao2015level}. Although faster, this approach often fails with asymmetrically shaped objects, as intersections between bounding spheres or cubes may not correspond to intersections among the actual object meshes, as shown in Figure \ref{fig:intersection} (b, c). This can lead to excessive penalization of intersections, causing the network to push objects apart and create unnecessary gaps.

\textbf{\textit{Intersection Loss.}} 
To overcome these challenges, we propose an SDF-based approach to handle both object-object intersections and object-container extrusion. The signed distance field (SDF) provides the shortest distance from a point to the surface, with a negative sign indicating the point is inside and a positive sign indicating it is outside the object. Specifically, we pre-compute the SDF  of each mesh (using \cite{Wang-2022-dualocnn}) at fixed query points within the container $\mathcal{C}$ (a few milliseconds on GPU) and deform the SDFs with every update of the learned rotation and translation parameters. Rather than recalculating the SDF for each rigid transformation, we warp the previous SDF using a linear transformation to match the new configuration, as shown in Figure~\ref{fig:intersection} (d). The degree of intersection, thus, can be determined by identifying points where more than one object has a negative SDF value, as marked in red in Figure \ref{fig:intersection} (d).

\begin{figure}[!ht]
    \centering
    \includegraphics[width=\linewidth]{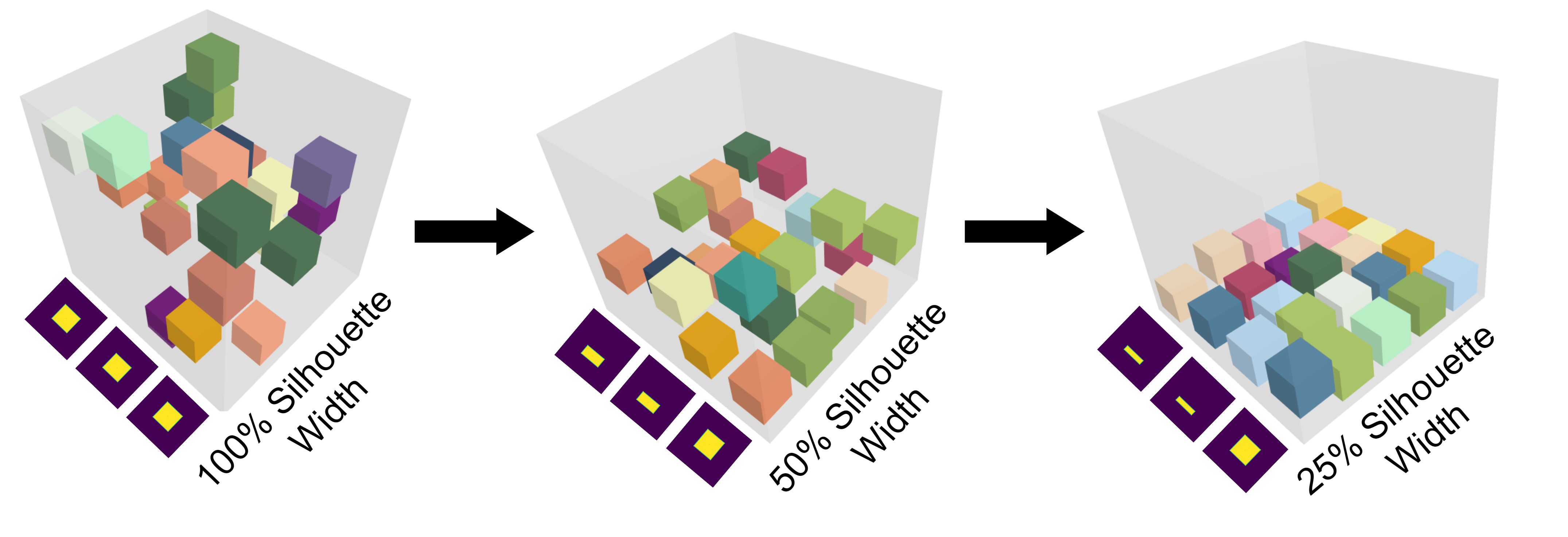}
    \caption{Illustration of packing of 24 cubes, each of dimensions $1\times1\times1$ cubic units, into a container of dimensions $6\times6 \times6$ cubic units, with varying silhouette widths to simulate physically correct packing for $N < N_{max}$.}
    \label{fig:strip}
    \vspace{-0.25cm}
\end{figure}

Let the SDF of the object $O_{i}$ at a point $\mathbf{p}$ after a rigid transformation (updated rotation and translation) has been applied to $O_{i}$ be denoted by $\widetilde{S}_{O_{i}}(\mathbf{p})$. One could simply consider the amount of intersection to be proportional to the number of objects under intersection at any point within the container, \emph{i.e.}, $D_{is}(\mathbf{p}) = \big| \{i \in \{1,2,...N\} | \widetilde{S}_{O_{i}}(\mathbf{p}) < 0 \} \big|$. However, we observed that considering the SDF value instead of just the count provides smoother guidance to resolve the intersections. Therefore, we define the degree of intersection $D_{is}(\mathbf{p})$ as the sum of the (negative of) SDFs of the objects that contain point $\mathbf{p}$ within them at every optimization step, such that,
\begin{equation}
    \centering
    D_{is}(\mathbf{p}) = \sum_{\{\forall O_{i} | \widetilde{S}_{O_{i}}(\mathbf{p}) < 0\}} - \widetilde{S}_{O_{i}}(\mathbf{p})
\end{equation}

Here, $D_{is}(\mathbf{p}) > 1$ indicates that there is an intersection of at least two objects at the point, and $D_{is}(\mathbf{p}) = 0$ represents that the point $\mathbf{p}$ is not inside any object. Considering the set of query points $\mathcal{C}_{p}$ inside the container $\mathcal{C}$, the intersection loss is defined as per Equation \ref{eq:inter_loss}.
\begin{equation}
    \centering
    \mathcal{L}_{is} = \sum_{\mathbf{p} \in \mathcal{C}_{p}} D_{is}(\mathbf{p})
    \label{eq:inter_loss}
\end{equation}


\textbf{\textit{Container Extrusion Loss.}} Let $V_{i}$ be the set of vertices of the triangular mesh corresponding to the object $O_{i}$ and $S_{\mathcal{C}}$ be the SDF of the container $\mathcal{C}$. We define the container extrusion loss as per Equation \ref{eq:ext_loss}
\begin{equation}
    \centering
    \mathcal{L}_{ext} = \sum_{i=1}^{N} \sum_{\mathbf{v} \in V_{i}} \text{max}(-\epsilon, S_{\mathcal{C}}(\mathbf{v})) 
    \label{eq:ext_loss}
\end{equation} 
Here, $\epsilon$, the distance between a vertex and the container's boundary, creates a ``buffer zone" around the container boundary to control when the extrusion loss begins to be active. While a higher value of $\epsilon$ hinders the movement of objects inside the container, a very small value leads to portions of objects suddenly extruding out of the container boundary, leading to an unstable optimization. We find $\epsilon = 0.01$ to be a reasonable value that avoids penalizing objects that are slightly inside the container but still close to the boundary. We observed that in some cases, silhouette loss was just enough to contain the objects with the container, especially for simple shapes. However, for more irregular shapes with asymmetric and skewed spread, extrusion loss proved to be helpful.

\begin{figure}[t]
    \centering
    \includegraphics[width=\linewidth]{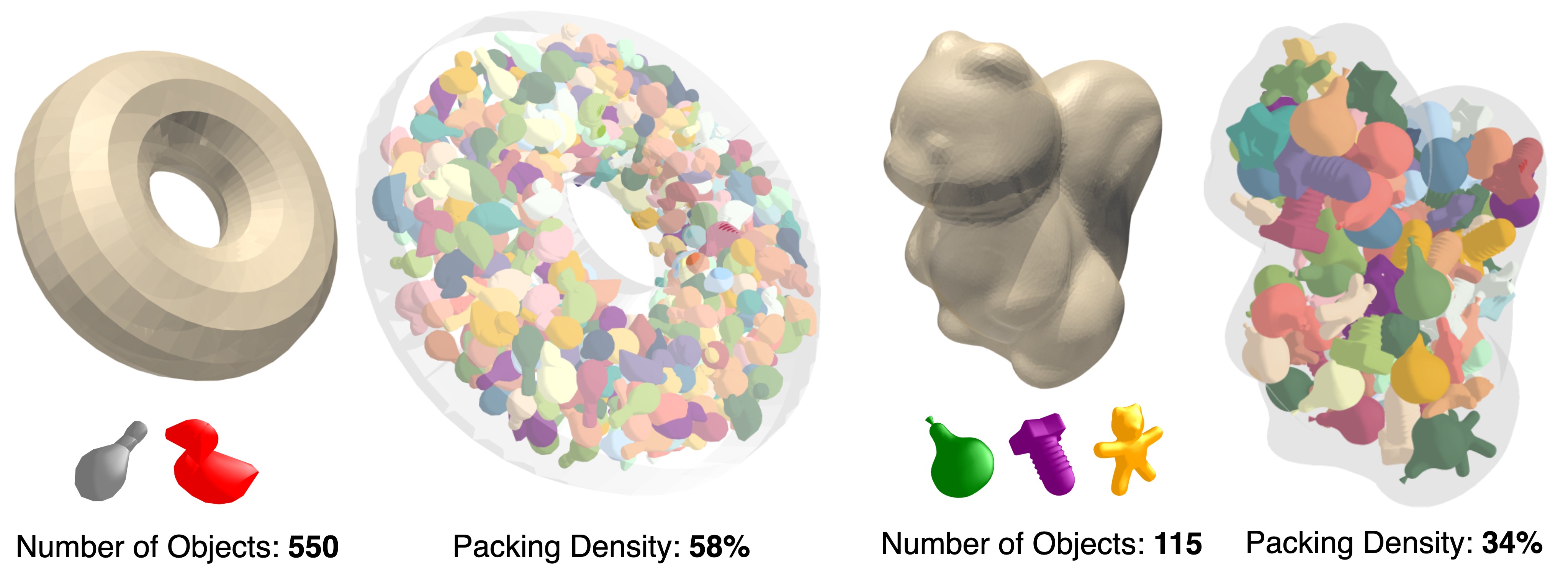}
    \caption{Illustration of packing arbitrarily shaped objects into arbitrarily shaped containers. The initial containers, the set of object shapes to be packed, and the final packed configuration are shown in the figure.}
    \label{fig:custom_container}
\end{figure}

\section{Experiments}
\label{sec:exp}
In this section, we showcase the results obtained through the RASP framework for irregular object packing drawing parallels with 3D Anamorphic art. Furthermore, we extend RASP to applications like part assembly and creating artistic illustrations. 

\subsection{Irregular Object Packing}
For 3D object packing, we demonstrate the results majorly on the objects from the IR-BPP dataset \cite{zhao2023learning}, which contains objects from four categories: \textit{General}, \textit{Kitchen}, \textit{ABC}, and \textit{Block Out}. A few other illustrations also include objects from the \textit{VOLMAP} dataset \cite{volmap2023} and \textit{SilNet} dataset \cite{zisserman2017silnet}. 

Following the existing literature of 3D object packing, we define the packing efficiency/density $(\rho)$, as per Equation \ref{eq:6}.
\begin{equation}
    \centering
    \rho = \sum_{O_{i}\in\mathcal{C}}\frac{|O_{i}|}{|\mathcal{C}|}
    \label{eq:6}
\end{equation}
Here, $|O_{i}|$ and $|\mathcal{C}|$ are the volume of the $i^{th}$ object and container volume, respectively.

Figure \ref{fig:container}, compares the packing density of irregular 3D objects from different categories into a typical cuboidal container. Interestingly, RASP does not require any regularity in object shapes and can handle arbitrary shapes of different sizes, achieving reasonably good packing configurations.

\begin{figure*}[t]
    \centering
    \includegraphics[width=\textwidth]{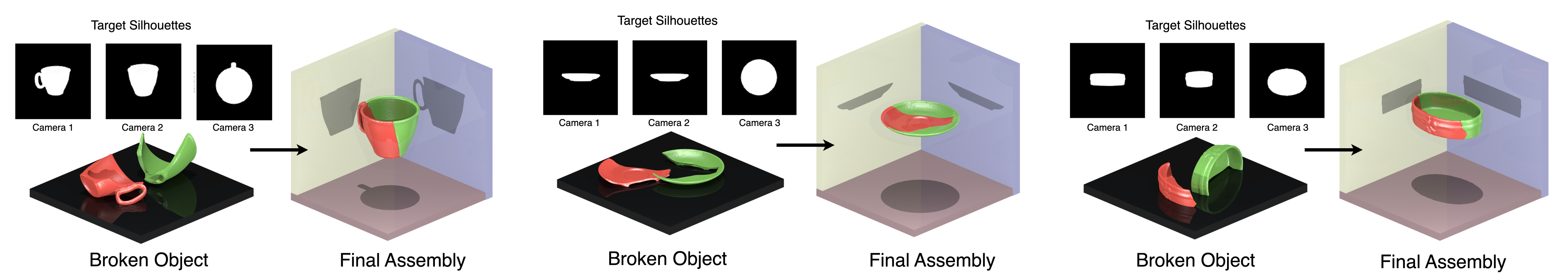}
    \caption{Extending RASP to perform part reassembly of broken objects. The figure showcases the target silhouettes and broken parts reassembled using RASP.}
    \label{fig:part_asm}
    \vspace{-0.5cm}
\end{figure*}

Deciding the number of objects that can be maximally packed within a container is crucial for optimal packing. To start with, we decide the initial number of objects as $N_{init}$  based on the average volume of the objects and the container volume, such that $N_{init} = \frac{|\mathcal{C}|}{|O|_{avg}}$. Consider $N_{max}$ as an unknown upper bound, \emph{i.e.},  the maximum a physical container can accommodate. If $N_{init} < N_{max}$, we draw another set of $N_{k}$ objects from the pool (if available) and again optimize the arrangement with $N_{init} + N_{k}$ based on the target silhouettes of projection. In case this exceeds $N_{max}$, the remaining objects either start extruding out of the container or intersecting heavily within the container depending on the predominance of $\mathcal{L}_{is}$ or $\mathcal{L}_{ext}$, respectively. All the results correspond to $N = N_{max}$ objects that can maximally fit inside the container with \textit{no intersection}, a condition essential for real-world applicability. Moreover, the objects are \textit{randomly initialized} within the container before optimization. We discuss the effect of different initialization strategies in the supplementary.

Due to silhouette or shadow-guided optimization, the resulting arrangement can sometimes be physically inaccurate. As shown in Figure \ref{fig:strip}, when the number of objects $N$ is less than the maximum capacity $N_{max}$, the arrangement often floats within the container rather than settling at the bottom. While such effects are inherent to this type of optimization-based framework, we demonstrate that merely adjusting the width of the silhouettes would effectively constrain the objects to rest at the bottom of the container in Figure \ref{fig:strip} progressively.

\textbf{\textit{Arbitrarily Shaped Container.}}
Owing to its design to derive insights from the projected silhouettes, RASP can also accommodate arbitrarily shaped containers, as shown in Figure \ref{fig:teaser} (b) and Figure \ref{fig:custom_container}. Similar to multi-view shape optimization, the target shadows are essentially the silhouettes of the container from $5$ different views. Figure \ref{fig:custom_container} depicts the packing on donut-shaped and squirrel-shaped containers.

\textbf{\textit{Effect of loss terms.}} Figure \ref{fig:ablation} (top) shows the convergence of each of the loss terms for illustrations in Figure \ref{fig:container} and Figure \ref{fig:custom_container}. The intersection density for all the results is zero, offering a multi-view consistent, intersection-free/non-overlapping configuration. Figure \ref{fig:ablation} (bottom) also illustrates the impact of different loss terms in addition to observations in Figure \ref{fig:intersection} for $\mathcal{L}_{sil}$ vs  $\mathcal{L}_{sil} + \mathcal{L}_{in}$. Overall, we find that extrusion term $\mathcal{L}_{ext}$ acts as a regularizer, preventing objects from drifting too far apart to avoid intersections. While the intersection remains zero without $\mathcal{L}_{ext}$, its inclusion improves the rendered silhouette by reducing $\mathcal{L}_{sil}$. Notably, objects are positioned closer together when $\mathcal{L}_{ext}$ is applied. 

\begin{figure*}[t]
    \centering
    \includegraphics[width=\textwidth]{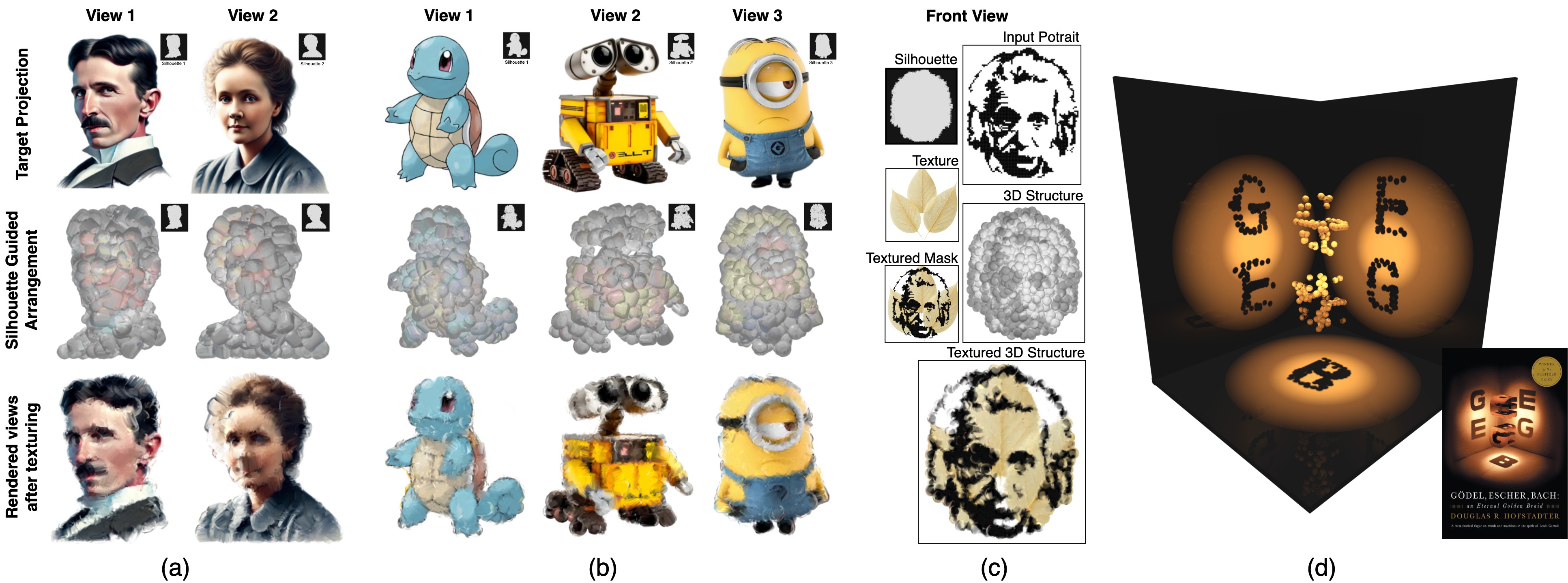}
    \caption{(a) \& (b) use objects from the Kitchen dataset to generate 2-view and 3-view 3D anamorphic art. (c) Multi-view pixelated portraits generated using RASP with artistic texture applied. (d) RASP recreates the famous cover page of the book Gödel, Escher, Bach by Douglas Hofstadter. The 3D and dynamic visualization of these and more related results are provided in the supplementary. }
    \label{fig:collage}
    \vspace{-0.5cm}
\end{figure*}

\textbf{\textit{Comparison with existing methods.}}
Out of several relevant works on 3D object packing (as described in Section \ref{sec:related_work}), the work by Ma \emph{et al.} \cite{ma2018packing} is the closest to our optimization setup (not guided by shadows) whose setup is different from RASP since it involves swapping, replacement with new objects, and object enlargement - which we believe is not practical in a real-world setup where we cannot alter the object dimensions. However, since their implementation is unavailable online, we choose to quantitatively compare the average packing efficiency across different objects.  Although different from ours, we also compared our average packing efficiency of a physically inspired reinforcement learning-based online packing method by Zhao \emph{et al.} \cite{zhao2023learning} over their dataset to establish the efficacy of our optimization-based methods learning solely from 2D image guidance. Zhao \emph{et al.} \cite{zhao2023learning} look at only one or a few objects at a time, and packing efficiency is dependent on the sequence in which the objects arrive. Moreover, the object placement is guided by the dynamics and constraints of physics. In contrast, RASP - an \textit{offline} method, takes a more global standpoint by optimizing all $N \leq N_{max}$ objects at a time.  Overall, RASP obtains an average of $45\%$ occupancy over the four different categories of the IR-BPP dataset \cite{zhao2023learning} which is better than Ma \emph{et al.} \cite{ma2018packing} ($34\%$) and drops below that of Zhao \emph{et al.} \cite{zhao2023learning} ($51.9\%$) evaluated over the objects from online packing dataset. Moreover, Ma \emph{et al.} \cite{ma2018packing} bears an average optimization time of $40.55$ minutes while RASP achieves the same in $\sim15$ minutes.

For a qualitative comparison over samples from IR-BPP dataset, we adapt \cite{zhao2023learning} to align closest to our setting. Specifically, we generate $100$ random sequences of the same set (and same number) of objects from each IR-BPP dataset category (as in Figure \ref{fig:container}) and report the best configuration to compare with RASP in Figure. \ref{fig:qual_res}. Due to implementation constraints of \cite{zhao2023learning}, we could not assign different colors to objects or obtain multiple views of the packed arrangement of \cite{zhao2023learning}, and hence, we compared only the top view. RASP performs similar to or slightly better than the physics-aware method for the same object set.

\subsection{3D Part Assembly}

Part assembly using RASP also shares similarities with multi-view geometry optimization. However, instead of optimizing a single shape, RASP learns the rigid transformations of different parts of a single shape to obtain a 3D consistent arrangement across all the views. For demonstrating part assembly, we used the \textit{Fantastic Breaks} dataset \cite{lamb2023fantastic} that consists of paired 3D scans of real-world broken objects and their complete counterparts.  Figure~\ref{fig:part_asm} demonstrates some qualitative results on reassembling broken objects. Notably, it does so solely via silhouette guidance without the need for any explicit 3D ground truth supervision.



\subsection{Multi-view Anamorphic Art}
We also leverage RASP to reinterpret and construct different forms of multi-view anamorphic art. Thus far, we have seen that silhouettes or shadows do provide important cues for 3D arrangement. However, these binary images alone cannot create interesting artistic illustrations. Therefore, we also seek guidance from colored or textured images. Take, for example, the portrait of Nikola Tesla shown in Figure \ref{fig:1} (a). One approach would be to search for suitable objects and apply heuristic methods to arrange them in a way that recreates the portrait’s appearance. Alternatively, one could collect and randomly arrange objects (like those in Figure \ref{fig:1} (c)) within the outline of the silhouette, then paint the arrangement to revive the essence of the portrait. Following the latter approach, RASP uses a two-stage optimization procedure, where first, it arranges the discrete objects in a 3D space through silhouette matching across a set of views and later performs rendering-based texture optimization (integrated as an add-on) over the finalized arrangement to match the target textures. Figure~\ref{fig:collage} (a) illustrates how this strategy allows RASP to recreate multi-view portraits of Nikola Tesla and Marie Curie across two non-orthogonal views that are $120^{\circ}$ apart using objects from the \textit{Kitchen} dataset. 
Furthermore, it also generates consistent ensembles that are meaningful across three non-orthogonal views, as shown in Figure \ref{fig:collage} (b), giving a visually plausible appearance of Pokémon, Wall-e, and Minion. In figure \ref{fig:collage} (c) we obtain 3D arrangements using binary images and then apply artistic textures to create unique 3D illustrations. The resemblance of the textured arrangement and the associated rendered images highlights the creative potential of RASP. In Figure \ref{fig:collage} (d) we recreate the famous artistic cover page of the Book by Douglas Hofstadter featuring blocks casting shadows of the first letters of artists -- Gödel, Escher, and Bach. Overall, RASP is a versatile optimization pipeline stressing on the fact that shadows do provide limited yet useful cues for 3D understanding and artistic exploration.


\section{Conclusion}
\label{sec:conclusion}

We introduce RASP, a differentiable rendering-based optimization framework for irregular 3D object packing, part assembly, and recreating artistic illustrations, taking guidance from images, whether in the form of binary silhouettes, textured RGB images, or simple portrait sketches. The current offline strategy does not account for physical dynamics, such as the influence of gravity on object placement. Additionally, RASP struggles with multi-part (more than 2 or 3 parts) part assembly, particularly symmetrical and/or identical parts. An interesting potential extension of this work would be to incorporate physics-based guidance for the packing to be more physically consistent. We believe that there are plenty of untapped capabilities in shadows that drive 3D understanding, and RASP presents a few of them. We anticipate that this work would attract researchers from different domains to use shadows or silhouettes for applications like partitioning and reconfiguration of complex 3D objects into simpler forms, multi-part part assembly, dynamic visual arts, CAD design, and handling non-rigid deformations.

\noindent\textbf{Acknowledgments.} This work is supported by the Prime Minister Research Fellowship (PMRF) grant and the Jibaben Patel Chair in Artificial Intelligence. We also thank Prajwal Singh, IIT Gandhinagar, for his valuable inputs.



{
    \small
    \bibliographystyle{ieeenat_fullname}
    \bibliography{main}
}


\end{document}